\documentclass{aa}  

\usepackage{graphicx}
\usepackage{txfonts}
\usepackage{threeparttable}
\usepackage{subcaption}
\usepackage{multirow}
\usepackage{makecell}
\usepackage{hyperref}
\hypersetup{
	colorlinks=true,
	linkcolor=blue,
	filecolor=blue,      
	urlcolor=blue,
	citecolor=blue,
    breaklinks=true
}
\usepackage{siunitx}
\usepackage{booktabs}

\begin{document}

   \title{Pre-perihelion radio observations of comet 12P/Pons-Brooks with Tianma radio telescope}

   \author{Juncen Li
          \inst{1,2,3}
    \and
          Xian Shi
          \inst{3}
    \and
          Jianchun Shi
          \inst{3}
    \and
          Yuehua Ma
          \inst{1,2}
    \and
          Bin Yang
          \inst{4}
    \and
          Jian-Yang Li
          \inst{5}
    \and
          Zexi Xing
          \inst{3}
    \and
          Man-To Hui
          \inst{3}
    \and
          Zhen Wang
          \inst{6}
    \and
          Ruining Zhao
          \inst{7}
          }

   \institute{Key Laboratory of Planetary Sciences, Purple Mountain Observatory, Chinese Academy of Sciences, Nanjing 210023, China
   \email{yhma@pmo.ac.cn}
         \and
              School of Astronomy and Space Science, University of Science and Technology of China, Hefei 230026, China
        \and
            Shanghai Astronomical Observatory, Chinese Academy of Sciences, 80 Nandan Road, 200030 Shanghai, People's Republic of China
            \email{shi@shao.ac.cn}
        \and
             Instituto de Estudios Astrofísicos, Facultad de Ingeniería y Ciencias, Universidad Diego Portales, Santiago, Chile
        \and
             Planetary Environmental and Astrobiological Research Laboratory (PEARL), School of Atmospheric Sciences, Sun Yat-sen University, Zhuhai, China
        \and
             Key Laboratory of Radio Astronomy, Xinjiang Astronomical Observatory, Chinese Academy of Sciences, Urumqi, Xinjiang 830011, China
        \and
             Key Laboratory of Optical Astronomy, National Astronomical Observatories, Chinese Academy of Sciences, Beijing 100101, China
             }

   \date{Received April 2, 2025; accepted August 5, 2025}
 
  \abstract
   {The multiple outburst events of comet 12P/Pons-Brooks during its 2024 apparition offer a unique window into highly-active volatile releasing processes not observable during quiescent periods.
   Radio observations are capable of measuring specific cometary gas species that offer valuable insights into the chemical composition and activity mechanism of comets.}
   {We performed radio observations of comet 12P/Pons-Brooks with the Tianma-65m radio telescope, targeting the OH and NH$_3$ inversion lines at 18-cm and 1.3-cm, respectively. By monitoring 12P at different heliocentric distances on its inbound journey, we aim to provide insights into the comet's volatile composition and outburst behavior.}
   {Four observations were carried out between December 2023 and March 2024 when the comet was approaching the Sun from 2.22 AU to 1.18 AU. We conducted 18-cm OH lines observations on 4 single days using the cryogenically cooled receiver system of the telescope to derive $\rm H_{2}O$ production rate. During 12P's outburst on December 14, we also conducted observations targeting the $\rm NH_{3}$ emission.}
   {OH 18-cm lines were clearly detected with a signal-to-noise ratio of $\sim$4$\sigma$ (peak intensity), yielding estimates on corresponding water production rates of approximately $10^{29}$ $\rm molec.\cdot s^{-1}$. A tentative detection of $\rm NH_{3}$ was made at the $\sim$$3\sigma$ level during the outburst phase, 
   but the detection needs to be further verified.}
   {Our observations provide information on the outgassing behavior of 12P/Pons-Brooks during its 2024 apparition. 
   The water production rate of 12P, derived from the 18-cm OH lines is consistent with measurements obtained in other works.
   The possible detection of $\rm NH_{3}$ during an outburst suggests possible connections between subsurface volatile reservoir and the outburst mechanism. 
   These results could further our understanding of the composition and activity of Halley-type comets.}

   \keywords{astrochemistry --
                comets: individual: 12P/Pons-Brooks --
                radio lines: general
               }

   \maketitle

\section{Introduction} 
\label{sec1}

Comets, as primitive remnants of the early solar system, serve as crucial probes to understand the physical and chemical conditions that prevailed during the formation of our planetary system \citep{2009oeec.book....5H}.
These "cosmic fossils" have preserved pristine materials from the proto-solar nebula, including ices, organics, and dust, largely unaltered since their formation about 4.6 billion years ago \citep{2011ARA&A..49..471M}. 
The study of cometary composition and activity not only provides insights into the chemical inventory and temperature conditions in the early solar system, but also helps trace the origin and evolution of volatile elements on Earth \citep{2019ARA&A..57..113A}.

When comets approach the planet region of the solar system, solar heating triggers the sublimation of their volatile components, primarily water ice, leading to the development of a coma and characteristic tails. 
This process, known as outgassing, releases various molecular species that can be observed through different techniques. 
Radio astronomy has emerged as a powerful technique for investigating cometary molecular species through their rotational transitions \citep{2004come.book..391B, 2017RSPTA.37560252B}.

Comet 12P/Pons-Brooks (hereafter 12P), discovered independently by Jean-Louis Pons in 1812 and William Brooks in 1883, is a Halley-type comet (HTC) with an orbital period of 71.3 years. 
Numerical integration suggested that the comet might not have experienced significant gravitational perturbations in the past $\sim$1000 years, and had retained a relatively stable orbit \citep{2020arXiv201215583M}. 
The most intriguing characteristic of 12P is its recurring intensive outbursts, as recorded during each of its apparitions since 1883 \citep{2003come.book.....K,2009come.book.....K,10.1093/mnras/115.2.190,https://doi.org/10.1002/asna.18841070902}. 
Sublimation of super-volatile ices might cause such a phenomenon, but the detailed mechanism is still unclear.
The OH 18-cm lines, resulting from the $\Lambda$-doubling transitions in the ground state of the OH radical, have been widely used as a proxy to determine water production rates in comets \citep{2002A&A...393.1053C}. 
These lines are particularly valuable because they are observable from the ground and provide reliable measurements of the water production rate, which is crucial for understanding cometary activity. 
The OH radicals are primarily produced through water photodissociation, with a nearly 90\% branching ratio \citep{1992Ap&SS.195....1H}.

Ammonia ($\rm NH_{3}$) is another important volatile species in comets, typically present at an abundance of $\sim$0.1-3.6$\%$ relative to water \citep{2023A&A...677A.157D}. 
Direct detection of $\rm NH_3$ in comets is relatively rare due to its short photodissociation lifetime, small abundance, and weak intrinsic line strengths. 
Observations of the inversion lines at 1.3 cm wavelength are especially scarce. 
These lines have been detected in comets C/1983 H1 (IRAS-Araki-Alcock) \citep{1983A&A...125L..19A}, C/1996 B2 (Hyakutake) \citep{1996AAS...188.6212P}, C/1995 O1 (Hale-Bopp) \citep{1997A&A...325L...5B, 1999ApJ...520..895H}, and not been detected in comets C/2001 A2 (LINEAR), C/2001 Q4 (NEAT), C/2002 T7 (LINEAR), 153P/Ikeya-Zhang \citep{2002ESASP.500..697B,2005A&A...439..777H}.
Among these results, the (3,3) transition is the most frequently detected and the strongest line among the five transitions.
In the infrared domain, particularly around 3 $\mu \rm m$, numerous $\rm NH_{3}$ detections have been reported despite the challenges posed by relatively weak line strengths and spectral blending with other molecular species \citep{2016Icar..278..301D}.
The fundamental rotational transition $(1,0)$$\rightarrow$$(0,0)$ at 572 GHz is only accessible from space and valuable measurements were obtained by the $Odin$ and $Herschel$ satellites \citep{2012A&A...539A..68B}. This same transition was crucial for monitoring $\rm NH_{3}$ in comet 67P/C-G using the MIRO instrument during the Rosetta mission \citep{2019A&A...630A..19B}.
However, during outbursts, enhanced production of $\rm NH_{3}$ may provide opportunities for ground-based detections.

In 2020, 12P was recovered at a heliocentric distance of 11.89 AU, as it was making the latest in-bound journey to the inner solar system \citep{2020RNAAS...4..101Y}. 
During its current apparition of 2024, the comet experienced several major and minor outbursts \citep{2025AJ....169..338J}.
Table.~\ref{tab2} showed the outbursts of 12P and the changes of brightness in 2023-2024, where the most prominent outbursts in July and November are readily recognizable. 
After the huge outburst that brightened the comet by 5 mag on July 19th, 2023, the activity of 12P has remained at a relatively high level with an outburst every few weeks.
Following the initial outburst, sustained brightening was observed, peaking around October 5, 2023 and November 14, 2023.
12P reached perihelion at a distance of 0.78 AU from the Sun on April 21, 2024 and became a naked-eye comet, whose activity grew stronger.

      \begin{table}
      \centering
      \caption{Pre-perihelion outbursts of comet 12P/Pons-Brooks.}
         \label{tab2}
         \setlength{\tabcolsep}{2mm}
         \begin{tabular}{ccc}
            \toprule
            Time (UT) & $\Delta m$ (mag) & Reference \\
            \midrule
            2023/07/20.82 & $\sim$ -5 & \citet{2023ATel16194....1M} \\
            2023/09/04.3 & $\sim$ -0.34 & \citet{2023ATel16229....1U} \\
            2023/09/25.15 & $\sim$ -0.87 &\citet{2023ATel16254....1K} \\
            2023/10/04.75 & $\sim$ -5 &\citet{2023ATel16270....1U} \\
            2023/10/22.93 to 25.85 & $\sim$ -0.63 &\citet{2023ATel16282....1J} \\
            2023/11/14.5 & $\sim$ -4.3 &\citet{2023ATel16338....1J} \\
            2023/12/12.6 to 14.6 & $\sim$ -1.5 &\citet{2024ATel16408....1J} \\
            2024/02/29.81 & $\sim$ -0.9 &\citet{2024ATel16498....1J} \\
            \bottomrule
         \end{tabular}
   \end{table}

In this paper, we present radio observations of comet 12P/Pons-Brooks using Tianma-65m radio telescope (TMRT), focusing on both OH 18-cm lines and 1.3-cm $\rm NH_{3}$ emission. 
Our observations span from December 2023 to March 2024, bracketing a significant outburst on December 14. These observations aim to:
\begin{enumerate}[1)]
\item Determine the water production rate and its temporal variations

\item Investigate the presence of $\rm NH_{3}$ during the outburst

\item Improve our understanding of the relationship between outbursts and volatile composition in Halley-type comets
\end{enumerate}
The paper is organized as follows: Section \ref{sec2} describes our observations and data reduction procedures. 
The results are presented in Section \ref{sec3}, followed by a discussion in Section \ref{sec4}. Our conclusions are summarized in Section \ref{sec5}.

\section{Observations and Data Reduction}
\label{sec2}

Observations on L-band and K-band centered at the 18-cm OH lines and the 1.3-cm $\rm NH_{3}$ lines, respectively, were carried out on four days, approximately 5 hours each day, between December 2023 and March 2024 using TMRT of Shanghai Astronomical Observatory in Shanghai, China \citep{2024AstTI...1..239L,2024AstTI...1..247D}. 
The frequencies were obtained from the JPL catalog\footnote{JPL catalog: https://spec.jpl.nasa.gov} \citep{1998JQSRT..60..883P}.
Observational parameters are provided in Table \ref{tab3}.
   
      \begin{table*}
      \centering
      \caption{Observational parameters of 12P/Pons-Brooks.}
         \label{tab3}
         \setlength{\tabcolsep}{4mm}
         \begin{threeparttable}
         \begin{tabular}{cccccccc}
            \toprule
            \makecell[c]{UT date \\ $[\rm yyyy/mm/dd.dd-dd.dd]$} & \makecell[c]{$\langle r_{\rm h} \rangle$$^{\rm a}$ \\ $[\rm AU]$} & \makecell[c]{$\langle \Delta \rangle$$^{\rm b}$ \\ $[\rm AU]$} & \makecell[c]{$\langle \alpha \rangle$$^{\rm c}$ \\ $[^\circ]$} & \makecell[c]{$ \langle v_{\rm r} \rangle $$^{\rm d}$ \\ $[\rm km \cdot s^{-1}]$} & \makecell[c]{Integ.$^{\rm e}$ \\ $\rm [min]$} & Line \\
            \midrule
            2023/12/14.17-14.45 & 2.22 & 2.46 & 23.57 & $-$19.01 & 234 & 1.3-cm $\rm NH_{3}$ lines \\
            \midrule
            2024/01/20.00-20.20 & 1.75 & 2.04 & 28.76 & $-$19.17 & 96 & 18-cm OH lines \\
            2024/03/02.23-02.41 & 1.19 & 1.68 & 35.40 & $-$8.55 & 99 & 18-cm OH lines \\
            2024/03/03.25-03.47 & 1.18 & 1.68 & 35.48 & $-$8.14 & 119 & 18-cm OH lines \\
            \bottomrule
         \end{tabular}

         \begin{tablenotes}
         \normalsize
         \item \textbf{Note:}
         $^{\rm a}$ Mean heliocentric distance during observation time; 
         $^{\rm b}$ Mean geocentric distance; 
         $^{\rm c}$ Mean solar phase angle (Sun–object–Earth); 
         $^{\rm d}$ Mean radial velocity; 
         $^{\rm e}$ On source integration time.
         The parameters are obtained from the JPL Horizons System (https://ssd.jpl.nasa.gov/horizons/) and the Minor Planet Ephemeris Service (MPEC; https://minorplanetcenter.net/iau/MPEph/MPEph.html)
         \end{tablenotes}
         \end{threeparttable}
      \end{table*}

The L-band receiver of TMRT employs a cryogenically cooled dual-polarization system \citep{2015AcASn..56..648L}. 
The system temperature is typically 20-30 K under good weather conditions \citep{2015AcASn..56...63W}.
The beam size is approximately 11.6 arcmin at 1.6 GHz, with an aperture efficiency of about 50\%. 
The receiver system includes a noise calibration unit and a temperature monitoring system \citep{2015ApJ...814....5Y,2016ApJ...824..136L}.
The K-band receiver features a dual-polarization cryogenic receiver system.
The typical system temperature is around 160 K in good weather.
The beam size is approximately 50 arcseconds at 24 GHz, with an aperture efficiency of about 50\%. The system employs a temperature-stabilized noise calibration unit and includes a water vapor radiometer for atmospheric calibration. 
The pointing accuracy is better than 5 arcseconds under normal conditions \citep{2017AcASn..58...37W}.
The DIBAS (Digital Backend System) of TMRT offers 29 different spectral modes, divided into single-window and multi-window configurations. 
The single-window modes include broadband options (Modes 1-3) with up to 1.5 GHz bandwidth and 16384 channels, and narrow-band options (Modes 4-19) achieving frequency resolutions as fine as 0.02 kHz. 
The multi-window modes (Modes 20-29) provide eight independent windows of 24 MHz or 16 MHz bandwidth that can be flexibly positioned within the receiving band, with frequency resolutions up to 0.24 kHz, enabling efficient simultaneous observations of multiple spectral lines with high spectral resolution across L to Q bands \citep{2015AcASn..56..648L}.

For K-band observations on 14 December 2023, we employed DIBAS mode 20, which provides eight spectral windows, each featuring 4096 channels across a 23.4 MHz bandwidth. 
This configuration yielded a velocity resolution of approximately 0.07 $\rm km \cdot s^{-1}$ at 24 GHz.
L-band observations utilized DIBAS mode 24, also with eight spectral windows, but with enhanced spectral resolution using 65536 channels per 23.4 MHz bandwidth window, achieving a velocity resolution of about 0.06 $\rm km \cdot s^{-1}$ at 1.6 GHz.
    
Single-point position-switching observations were conducted under favorable weather conditions.
The brightness temperature of the main beam ($T_{\rm MB}$) was derived from the antenna temperature ($T^{\ast}_{A}$) using $T_{\rm MB} = T^{\ast}_{A}/\eta_{B}$, where $\eta_{B} \sim 60\%$ represents the main beam efficiency. 
Prior to each observing session, pointing accuracy and system performance were verified using bright calibrator sources (e.g., W49N).
The conversion factor was 1.66 $\rm Jy\cdot K^{-1}$ under 50\% aperture efficiency from the antenna temperature $T_{A}^{*}$ to the flux density S \citep{2022ApJS..260...34Z}.

Data reduction and analysis were performed using the CLASS package\footnote{http://www.iram.fr/IRAMFR/GILDAS} within GILDAS software suite. 
The data processing pipeline included Doppler correction to account for the comet's radial velocity relative to the telescope during the observation period (as listed in Table~\ref{tab3}). 
The spectral analysis was confined to the right circular polarization data, as the left circular polarization measurements were compromised by much more Radio Frequency Interference (RFI) in the L-band.
Linear baselines were subtracted from each spectrum, excluding the emission line region. 
Individual spectra were then averaged with their rms noise levels as the weights. 
We verified that the baseline fitting procedure had negligible impact on the spectral line parameters within our narrow frequency windows.

\section{Results}
\label{sec3}

\begin{figure*}
   \centering
   \includegraphics[width=18cm]{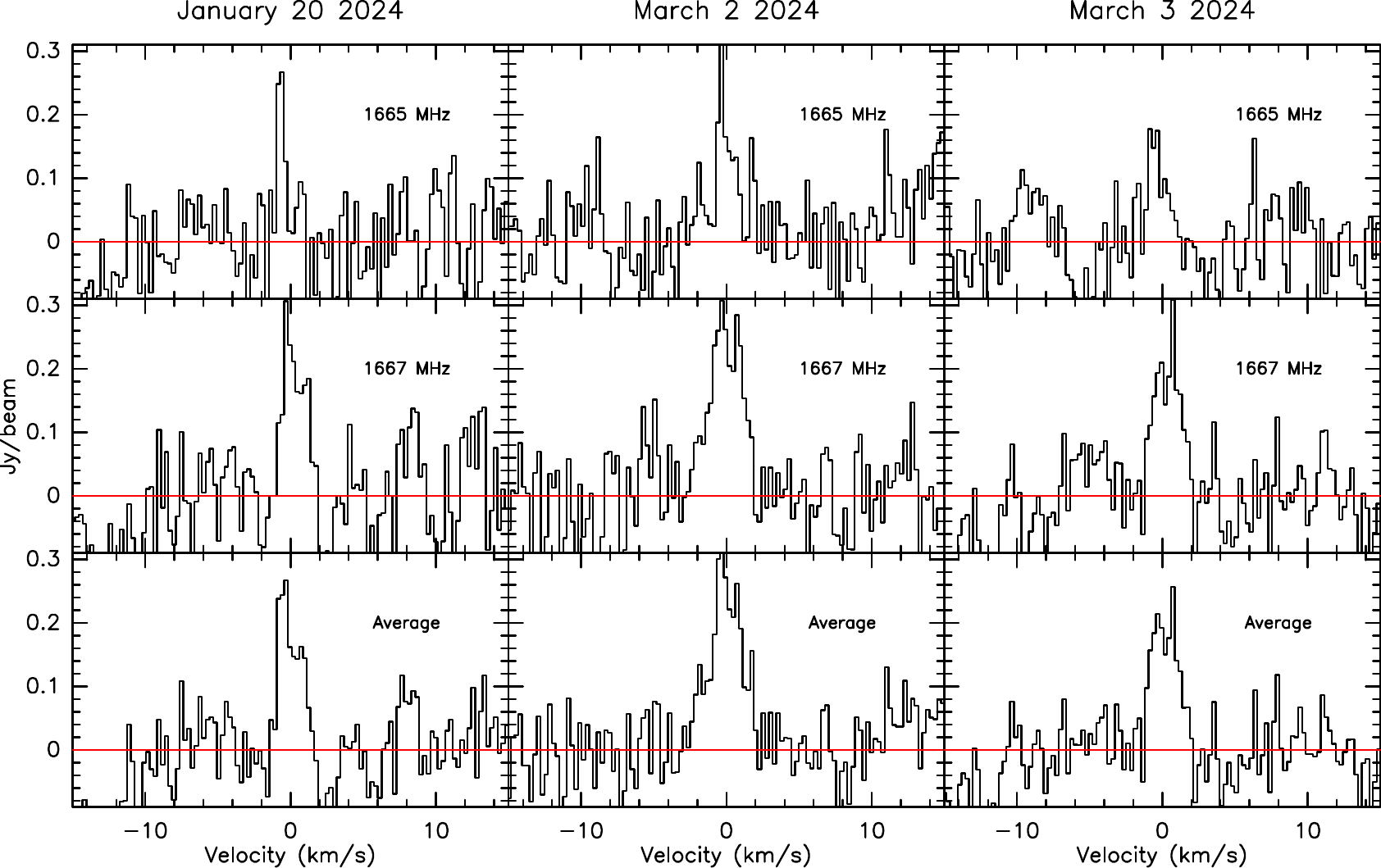}
      \caption{
      Averaged 18-cm OH lines of 12P/Pons/Brooks observed on three separate days: January 20 (left), March 2 (middle) and 3 (right), 2024.
      Top and middle row exhibited the 1665 and 1667 MHz lines, respectively. The bottom spectrum showed the weighted averages of the 1667 and 1665 MHz lines scaled to 1667 MHz, assuming the statistical ratio of 1.8.
      The vertical scale is the main beam brightness temperature and the horizontal scale is the Doppler velocity in the comet rest frame.
      The frequency resolution is 0.0014 MHz, corresponding to velocity resolution of 0.2575 $\rm km \cdot s^{-1}$ after smooth. 
      We listed the fitting results in Table~\ref{tab4} and \ref{tab5}. 
      }
         \label{fig:1}
   \end{figure*}

\begin{figure}
   \centering
   \includegraphics[width=\columnwidth, trim= 40 70 460 120, clip]{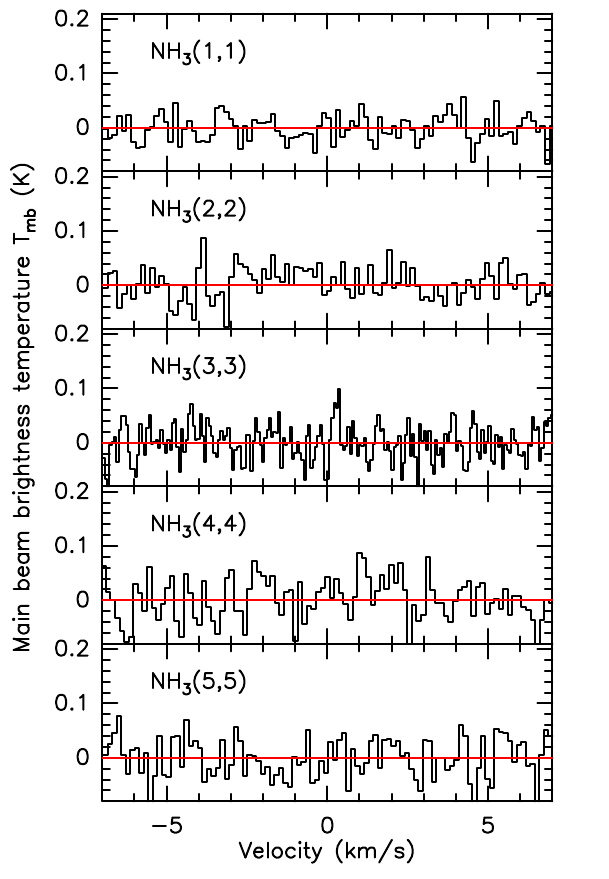}
      \caption{
      Averaged 1.3 cm $\rm NH_{3}$ lines of 12P/Pons-Brooks observed on December 14 2023.
      From top to bottom, they are {$\rm NH_{3}(1,1)$} to $(5,5)$. The frequency resolutions of {$\rm NH_{3} (3,3)$} was 0.057 MHz (about 0.0114 MHz for the other 4 lines), corresponding to velocity resolution of 0.0719 $\rm km \cdot s^{-1}$ (about 0.14 $\rm km \cdot s^{-1}$ for the other 4 lines) after smooth.
              }
         \label{fig:2}
   \end{figure}

The observed spectra of the 18-cm OH and 1.3-cm $\rm NH_{3}$ lines are shown in Fig.~\ref{fig:1} and \ref{fig:2}.
Spectra are aligned on the velocity scale.
For 18-cm OH lines, the main 1665 and 1667 MHz lines were clearly detected with a signal-to-noise ratio (SNR) above 4$\sigma$ (see details in Table~\ref{tab4}).
For the 5 $\rm NH_{3}$ lines at 1.3-cm, only (3, 3) was tentatively detected at an SNR of about 3$\sigma$ (see details in Table~\ref{tab6}). 

The expected statistical relative intensities of the 1667:1665 MHz lines are 9:5 (1.8) under local thermal equilibrium (LTE), which is in good agreement with the observations (1.6-2.1).
We analyzed the OH spectral line profiles with a Gaussian fit, and obtained the full width at half maximum (FWHM) and the Doppler shift relative to the reference frame of the cometary nucleus.
The integrated intensity was obtained from the velocity interval [$-2,2$] $\rm km \cdot s^{-1}$ ([$-2.5, +2.5$] $\rm km \cdot s^{-1}$ on March 2nd for the broader line profile), and the error was calculated from equation $\sqrt{n} \times {\rm RMS} \times dv$, where $dv$ is the channel width and $n$ is the number of channels that span the line.
The $3\sigma$ upper limit of the $\rm NH_{3}$ line intensities were obtained considering velocity interval [$-1.5$, $1.5$] $\rm km \cdot s^{-1}$.
The $\rm NH_{3}(3,3)$ line area was also obtained from a Gaussian fit.
Their values are summarized in Table~\ref{tab4} and ~\ref{tab6}.

      \begin{table*}
      \centering
      \caption{18-cm OH spectral characteristics of 12P-Pons-Brooks.}
         \label{tab4}
         \setlength{\tabcolsep}{3mm}
         \begin{threeparttable}
         \begin{tabular}{ccccccc}
            \toprule
            \makecell[c]{UT date \\ $[\rm yyyy/mm/dd]$} & lines & \makecell[c]{RMS$^{\rm a}$ \\ $\rm [Jy]$} & \makecell[c]{$ \rm FWHM $$^{\rm b}$ \\ $\rm [km \cdot s^{-1}]$} & \makecell[c]{$S^{\rm c}$ \\ $\rm [Jy \cdot km \cdot s^{-1}]$} & \makecell[c]{Doppler shift $^{\rm d}$ \\ $\rm [km \cdot s^{-1}]$} & \makecell[c]{$T_{\rm peak}^{\rm e}$ \\ $\rm [Jy]$} \\
            \midrule
            2024/01/20 & \makecell[c]{OH 1665 MHz \\ OH 1667 MHz \\ Average} & \makecell[c]{0.059 \\ 0.070 \\ 0.053} & \makecell[c]{$-$ \\ $-$ \\ $1.709 \pm 0.204$} & \makecell[c]{$0.200 \pm 0.060$ \\ $0.408 \pm 0.079$ \\ $0.392 \pm 0.054$} & \makecell[c]{$-$ \\ $-$ \\ $-0.114 \pm 0.111$} & \makecell[c]{0.266 \\ 0.306 \\ 0.267} \\
            \midrule
            2024/03/02 & \makecell[c]{OH 1665 MHz \\ OH 1667 MHz \\ Average} & \makecell[c]{0.057 \\ 0.067 \\ 0.045} & \makecell[c]{$-$ \\ $-$ \\ $2.587 \pm 0.229$} & \makecell[c]{$0.456 \pm 0.065$ \\ $0.740 \pm 0.076$ \\ $0.768 \pm 0.051$} & \makecell[c]{$-$ \\ $-$ \\ $-0.044 \pm 0.089$} & \makecell[c]{0.321 \\ 0.306 \\ 0.389} \\
            \midrule
            2024/03/03 & \makecell[c]{OH 1665 MHz \\ OH 1667 MHz \\ Average} & \makecell[c]{0.067 \\ 0.056 \\0.044} & \makecell[c]{$-$ \\ $-$ \\ $2.079 \pm 0.201$} & \makecell[c]{$0.253 \pm 0.068$ \\ $0.487 \pm 0.057$ \\ $0.479 \pm 0.045$} & \makecell[c]{$-$ \\ $-$ \\ $0.184 \pm 0.104$} & \makecell[c]{0.177 \\ 0.309 \\ 0.257} \\
            \bottomrule
         \end{tabular}

        \begin{tablenotes}
        \normalsize
        \item \textbf{Note:} 
        $^{\rm a}$ The $1\sigma$ noise of the base residuals in observed spectra ($S$ scale).
        $^{\rm b}$ The FWHM line width obtained from a Gaussian fit.
        $^{\rm c}$ The integrated flux density. 
        $^{\rm d}$ The velocity offset from a Gaussian fit.
        $^{\rm e}$ The peak flux.
        \end{tablenotes}
         \end{threeparttable}
      \end{table*}

\subsection{Hydroxyl data analysis}
\label{sec3.1}
\subsubsection{Outgassing pattern}
\label{sec3.1.1}

The OH line profiles observed on January 20 exhibited notable asymmetry, characterized by a distinct narrow feature at approximately -0.3 km/s in both the 1665 and the 1667 MHz transitions.
This asymmetric structure can be primarily attributed to the presence of a jet directed towards both Earth and the Sun, taking into account the small phase angle ($<30^\circ$) observed during this period.
The profiles became more symmetric as 12P moved closer to the Sun in March.

\begin{figure*}
    \centering
    \begin{subfigure}[b]{0.33\textwidth}
        \centering
        \includegraphics[width=\textwidth]{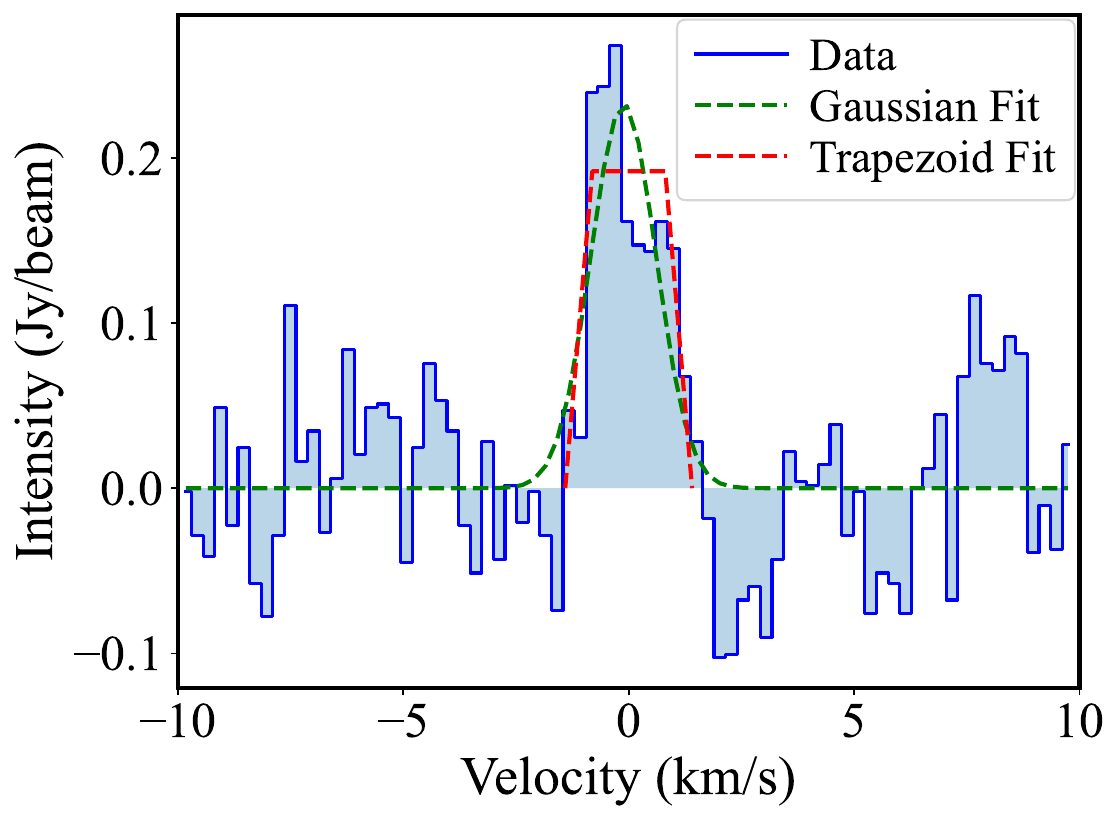}
        \caption{January 20}
    \end{subfigure}
    \hfill
    \begin{subfigure}[b]{0.33\textwidth}
        \centering
        \includegraphics[width=\textwidth]{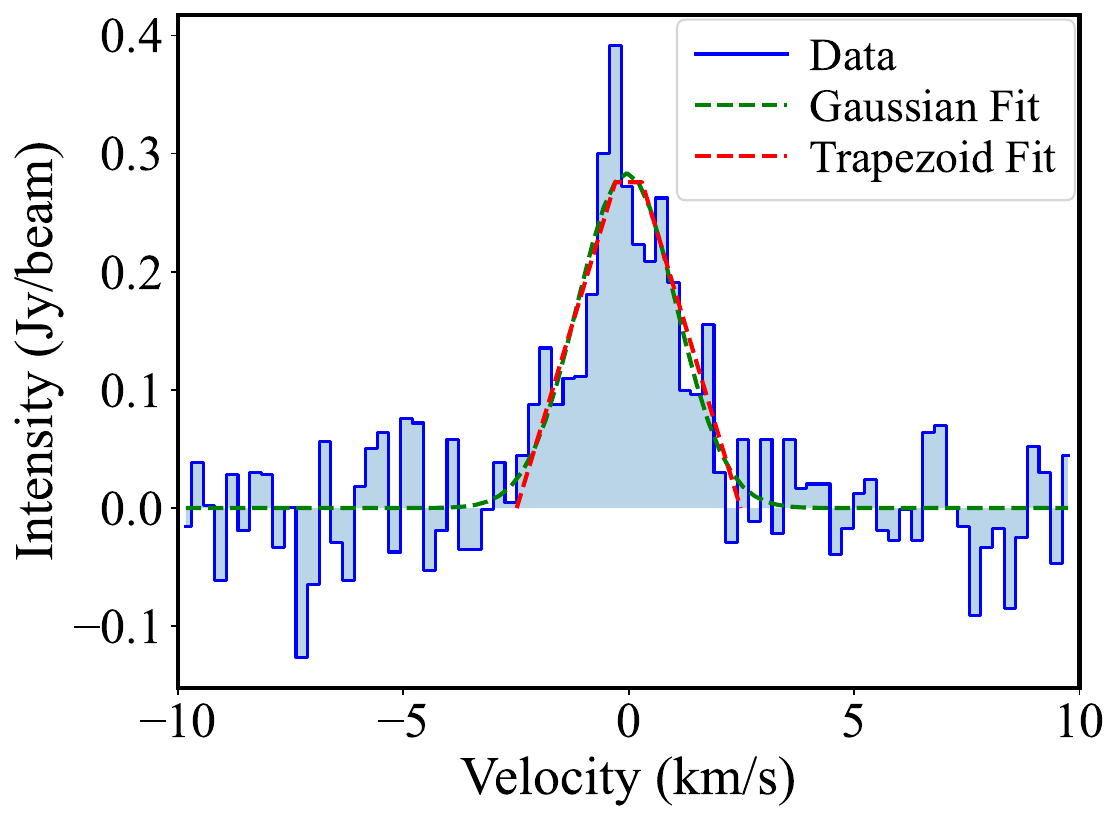}
        \caption{March 02}
    \end{subfigure}
    \hfill
    \begin{subfigure}[b]{0.33\textwidth}
        \centering
        \includegraphics[width=\textwidth]{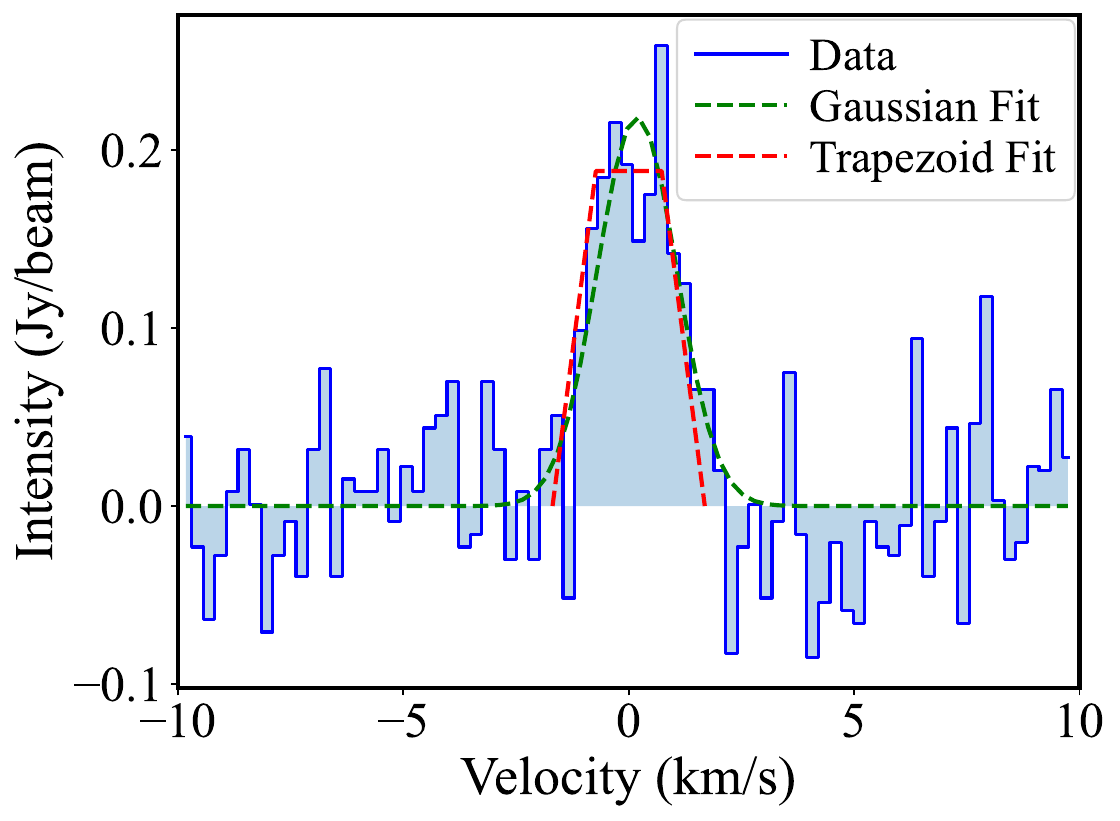}
        \caption{March 03}
    \end{subfigure}
    \caption{Averaged 18-cm OH lines (scaled to 1667 MHz) of 12P/Pons-Brooks obtained in three observations with both gaussian and trapezoid fit.}
        \label{fig:3}
    \end{figure*}

The observed line profiles were analyzed using the trapezoid model, which has been widely applied in radio observations of cometary OH lines (e.g. \citet{2002A&A...393.1053C}).
The OH velocity ($v_{\rm OH}$), comprising both the parent molecules outflow velocity ($v_{\rm p}$, e.g. $\rm H_{2}O$) and the ejection velocity ($v_{\rm e}$), was derived from fitting a symmetric trapezoidal profile to the spectral line (fig.~\ref{fig:3}). 
Under the assumption of monokinetic velocity distributions for both the parent molecule and OH ejection, the maximum radial velocity of OH along the line of sight is $v_{\rm p}$+$v_{\rm e}$ corresponding to the half-width of the trapezoid's lower base \citep{1990A&A...238..382B}.
The asymmetry in OH line profiles is a common characteristic in comets, primarily attributed to anisotropic outgassing patterns \citep{1984A&A...131..111B}. 
While this asymmetry has minimal impact on the lower base of the fitting trapezoid ($v_{\rm p}$+$v_{\rm e}$), it may lead to an underestimate of $v_{\rm p}$ \citep{1990A&A...238..382B}.
Although the FWHM derived from Gaussian fitting provides an alternative measure of velocity (Table~\ref{tab4}), determining $v_{\rm p}$ from the half-power width presents significant challenges for the daughter molecule spectral lines. 
This limitation was particularly evident in our January 20 observations, where the pronounced asymmetry in the line profile resulted in a notably larger discrepancy between the FWHM values obtained from Gaussian and trapezoidal fitting methods compared to other observation dates.
To address these uncertainties and facilitate the determination of $v_{\rm p}$, we adopted a standard value of 0.9 km/s for $v_{\rm e}$, which is consistent with the typical values observed in cometary studies \citep{1990A&A...238..382B}.
The derived values are presented in Table~\ref{tab5}.
$v_{\rm p}$ exhibited significant temporal variations throughout the observation period. 
Initially, $v_{\rm p}$ measured approximately 0.5 km/s when the comet was at a heliocentric distance of 1.75 AU. 
About two days following the February 29 mini-outburst, the measured velocity reached 1.4 km/s.
This represents a high value at heliocentric distances ($r_{\rm h}$) of 1.1-1.3 AU, considering the corresponding OH production rate \citep{2007A&A...467..729T}.
A lower value of 0.95 km/s was measured the next day.

\subsubsection{Production rate and evolution}

The OH 18-cm lines (1665/1667 MHz) in comets exhibit complex behavior due to the interaction between the OH radicals and the solar radiation field \citep{1981A&A....99..320D,1988ApJ...331.1058S}. 
The ground-state $^2\Pi_{3/2}$ $\Lambda$-doublet populations of OH are primarily determined by the solar UV pumping and collisional effects \citep{1981ApJ...246..354E,1982ApJ...258..864S}. 
The OH 18-cm lines act as a weak maser, with the population inversion of the $\Lambda$-doublet primarily controlled by the comet’s heliocentric velocity ($v_{\rm h}$) due to Doppler-shifted solar UV excitation.
In the inner coma regions of productive comets, collisional excitation dominates, equalizing the $\Lambda$-doublet level populations and quenching the maser inversion to zero \citep{1981A&A....99..320D,1990A&A...230..489G}.

The quenching radius ($r_{\rm q}$) defines the spherical region where maser quenching occurs, determined by the collision rate and then the number density of collisional partners (which is directly related to the water production rate here). \citep{1988ApJ...331.1058S,1991ASSL..167..149C}.
Inside $r_{\rm q}$, the OH radio lines are quenched, while outside $r_{\rm q}$, the lines maintain their normal emission characteristics. This effect creates a shell-like structure in the OH radio emission distribution, with reduced emission in the inner coma \citep{1998P&SS...46..569G}. 
The quenching phenomenon must be carefully considered when deriving OH production rates from radio observations, as it can lead to significant underestimation of the OH abundance if not properly accounted for in the analysis \citep{1989A&A...213..459C,2004come.book..391B}.

The number of unquenched OH radicals in the coma within the field of view (FOV) $\Gamma$ can be expressed as \citep{1985AJ.....90.1117S}:
\begin{equation}
    \Gamma_{\rm OH} ({\rm unquenched})=\frac{4\pi\Delta^{2}\nu S}{A_{\rm ul}kcT_{\rm bg}} \frac{8}{2F_{\rm u}+1}\bigg/\bigg[i+\frac{i+1}{2}\frac{h\nu}{kT_{\rm bg}}\bigg] \,,
\end{equation}
$A_{\rm ul}$ is the Einstein spontaneous coefficient of the
line at frequency $\nu$ ($A_{\rm ul}$ = $7.1 \times 10^{11} \rm s^{-1}$ and $7.7 \times 10^{11} \rm s^{-1}$ for the 1665 and 1667 MHz lines, respectively \citep{2013tra..book.....W}), $F_{\rm u}$ is the statistical weight of the upper level of the transition ($F_{\rm u}$ = 1 and 2 for the 1665 and 1667 MHz lines, respectively), $i$ is the inversion of the ground state $\Lambda$-doublet levels listed in Table~\ref{tab5}, $T_{\rm bg}$ is the back ground temperature, and $\Delta$ is the Earth-comet distance. 
Equation (1) was then simplified by \citet{1990A&A...238..382B} as:
\begin{equation}
    f\Gamma_{\rm OH}({\rm total})=2.33 \times 10^{34} \frac{\Delta^{2}S}{i T_{\rm bg}}\,,
\end{equation}
$f = \Gamma_{\rm OH}(\rm unquenched)/\Gamma_{\rm OH}(\rm total)$ is the fraction of the OH radicals at fluorescence equilibrium, which should be 1 for no collisional quenching effect when the FOV is much larger than the size of the OH coma.
Under our observation conditions, the FOV is 4-5 times larger than the scalelength of OH, $L_{\rm OH} = v_{\rm OH} \tau_{\rm OH}$, even when we assume an velocity of 1.4 km/s.
$\tau_{\rm OH}$ is the OH lifetime, $1.1 \times 10^{5}$ s at 1 AU \citep{1990A&A...230..489G,2002A&A...393.1053C} with $\tau_{\rm OH}=\tau_{\rm OH (1 AU)}\cdot r_{\rm h}^{2}$. 
For the calculation of $f$, the spatial distribution of OH radicals is described by the Haser model \citep{1957BSRSL..43..740H}:
\begin{equation}
n_{\rm OH}(r) = \frac{Q_{\rm p}}{4 \pi r^2 v_{\rm OH}} \cdot \frac{L_{\rm OH}}{L_{\rm OH} - L_{\rm p}} \cdot \left( e^{-r/L_{\rm OH}} - e^{-r/L_{\rm p}} \right)\,,
\end{equation}
$Q_{\rm p}$ is the production rate of water, the assumed primary parent molecule of OH. By adopting the branching ratio of OH, we could derive 12P’s water production rate as $Q_{\rm H_{2}O}=1.1\, Q_{\rm OH}$.
Similarly, the lifetime of $\rm H_2O$ $\tau_{\rm H_2O}=8.5 \times 10^{4}$ s at 1 AU was used to calculate $L_{\rm p}$ ($L_{\rm p}=v_{\rm p} \tau_{\rm H_{2}O}$).
Details of $v_{\rm p}$ and $v_{\rm OH}$ are provided in Sect.~\ref{sec3.1.1}.
The OH production rate $Q_{\rm OH}$ can be derived from $\Gamma_{\rm OH}$ by
\begin{equation}
    Q_{\rm OH}=\frac{\Gamma_{\rm OH}(\rm total)}{\tau_{\rm OH}}\,.
\end{equation}
For the determination of $Q_{\rm OH}$, we assumed an inversion $i = 0$ within the region $r<r_{\rm q}$.
Following \citet{1990A&A...230..489G} and \citet{1998P&SS...46..569G}, the quenching radius (in km) is related to the production rate and heliocentric distance as follows:
\begin{equation}
    r_{\rm q}=r_{\rm q}^{*}r_{\rm h}\sqrt{Q_{\rm OH}/10^{29}}\,\rm (km),
\end{equation}
where $r_{\rm q}^{*}=47000$ km.
An iterative method that simultaneously solves for both $Q_{\rm OH}$ and $r_{\rm q}$ was used. 
The calculation takes into account the Gaussian beam profile of the radio telescope and uses both spherical and cylindrical volume integrations to determine the number of OH radicals in different regions (e.g. \citet{1981A&A....99..320D,2023A&A...677A.157D}).

The OH production rates derived from our observations show a clear temporal evolution from January to March 2024. Starting at a relatively low rate of $Q_{\rm OH}$ = $(0.65 \pm 0.08) \times 10^{29} \rm molec. \cdot s^{-1}$ on January 20, the production increased significantly to reach a maximum of $Q_{\rm OH}$ = $(2.21 \pm 0.13) \times 10^{29} \rm molec. \cdot s^{-1}$ on March 2, followed by a decrease to $Q_{\rm OH}$ = $(1.39 \pm 0.11) \times 10^{29} \rm molec. \cdot s^{-1}$ on March 3 (listed in  Table~\ref{tab5}).
This variation represents nearly a factor of three increase in activity over approximately six weeks, followed by a $\sim$37\% decrease in just one day. 
The variation observed between March 2 and 3 is possibly related to the February 29 outburst.
The production rates calculated using the \citet{1988ApJ...331.1058S} inversion method are systematically higher by about 9-10\% compared to those derived using the \citet{1981A&A....99..320D} method, but both show the same temporal trend. 

      \begin{table*}
      \centering
      \caption[]{Averaged spectral characteristics of 18-cm OH lines and production rate in 12P-Pons-Brooks.}
         \label{tab5}
         \setlength{\tabcolsep}{2mm}
         \begin{threeparttable}
         \begin{tabular}{cccccccc}
            \toprule
            \makecell[c]{UT date \\ $[\rm yyyy/mm/dd]$} & \makecell[c]{$\langle \Dot{r_{\rm h} \rangle}$$^{\rm a}$ \\ $\rm [km \cdot s^{-1}]$} & $i$ (DE)$^{\rm b}$ & $i$ (SC)$^{\rm c}$ & \makecell[c]{$T_{\rm bg}$$^{\rm d}$ \\ $[\rm K]$} & \makecell[c]{$v_{\rm p}+v_{\rm e}$$^{\rm e}$ \\ $\rm [km \cdot s^{-1}]$} & \makecell[c]{$Q_{\rm OH}$ (DE) $^{\rm f}$ \\ $[\rm \times 10^{29} molec.\cdot s^{-1}]$} & \makecell[c]{$Q_{\rm OH}$ (SC) $^{\rm g}$ \\ $[\rm \times 10^{29} molec.\cdot s^{-1}]$} \\
            \midrule
            2024/01/20 & -22.81 & 0.49 & 0.44 & 3.9 & $1.42 \pm 0.14$ & $0.61 \pm 0.08$ & $0.68 \pm 0.09$ \\
            \noalign{\smallskip}
            2024/03/02 & -22.02 & 0.51 & 0.47 & 3.2 & $2.33 \pm 0.23$ & $2.11 \pm 0.13$ & $2.30 \pm 0.14$ \\
            \noalign{\smallskip}
            2024/03/03 & -21.91 & 0.51 & 0.47 & 3.2 & $1.86 \pm 0.19$ & $1.33 \pm 0.11$ & $1.45 \pm 0.12$ \\
            \bottomrule
         \end{tabular}
        \begin{tablenotes}
        \normalsize
        \item \textbf{Note:} 
        $^{\rm a}$ Mean radial velocity with respect to Sun.
        $^{\rm b}$ Maser inversion from \citet{1981A&A....99..320D}.
        $^{\rm c}$ Maser inversion from \citet{1988ApJ...331.1058S}.
        $^{\rm d}$ The background temperature at 1667 MHz was measured by interpolating the continuum maps at 408 MHz \citep{2015MNRAS.451.4311R,1974A&AS...13..359H} following the method mentioned in \citet{1981A&A....99..320D}.
        $^{\rm e}$ Half lower bases of the fitted trapezia.
        $^{\rm f}$ Production rate of OH calculated using the inversion from \citet{1981A&A....99..320D}.
        $^{\rm g}$ Production rate of OH calculated using the inversion from \citet{1988ApJ...331.1058S}.
        \end{tablenotes}
         \end{threeparttable}
      \end{table*}

      \begin{table*}
      \centering
      \caption[]{1.3-cm $\rm NH_{3}$ spectral characteristics of 12P-Pons-Brooks.}
         \label{tab6}
         \setlength{\tabcolsep}{4mm}
         \begin{threeparttable}
         \begin{tabular}{ccccccc}
            \toprule
            \makecell[c]{UT date \\ $[\rm yyyy/mm/dd]$} & lines & \makecell[c]{RMS$^{\rm a}$ \\ $\rm [K]$} & \makecell[c]{$ \rm FWHM $$^{\rm b}$ \\ $\rm [km \cdot s^{-1}]$} & \makecell[c]{$\int T_{\rm MB} dv$$^{\rm c}$ \\ $\rm [K \cdot km \cdot s^{-1}]$} & \makecell[c]{Doppler shift $^{\rm d}$ \\ $\rm [km \cdot s^{-1}]$} & \makecell[c]{$T_{\rm peak}^{\rm e}$ \\ $\rm [K]$} \\
            \midrule
            2023/12/14 & \makecell[c]{(1, 1) \\ (2, 2) \\ (3, 3) \\ (4, 4) \\ (5, 5)} & \makecell[c]{0.026 \\ 0.029 \\ 0.033 \\ 0.043 \\ 0.039} & \makecell[c]{$-$ \\ $-$ \\ $0.221 \pm 0.050$ \\ $-$ \\ $-$} & \makecell[c]{$<$0.051 \\ $<$0.058 \\ $<$0.045 / $0.022 \pm 0.006$ \\ $<$0.083 \\ $<$0.076} & \makecell[c]{$-$ \\ $-$ \\ $0.295 \pm 0.030$ \\ $-$ \\ $-$} & \makecell[c]{$-$ \\ $-$ \\ 0.103 \\ $-$ \\ $-$} \\
            \bottomrule
         \end{tabular}

        \begin{tablenotes}
        \normalsize
        \item \textbf{Note:} 
        $^{\rm a}$ The $1\sigma$ noise of the base residuals in observed spectra ($T_{\rm MB}$ scale).
        $^{\rm b}$ The FWHM line width obtained from a Gaussian fit.
        $^{\rm c}$ The integrated intensity or the $3\sigma$ upper limit of the $\rm NH_{3}$ line intensities were obtained from velocity interval of [$-1.5$, $1.5$] $\rm km \cdot s^{-1}$.
        $^{\rm d}$ The velocity offset from a Gaussian fit.
        $^{\rm e}$ The peak main beam temperature.
        \end{tablenotes}
         \end{threeparttable}
      \end{table*}

\subsection{Ammonia data analysis}

Among the 5 metastable ammonia lines at 1.3-cm, only the strongest (3, 3) were detected.
As discussed in \citet{2023A&A...677A.157D}, only the (2, 2) line competes in intensity with the (3, 3) line,  taking into account the distribution of the hyperfine components.
The narrow line width of the marginal $\rm NH_3(3,3)$  ($\sim$0.11 km/s) and the significant red Doppler shift ($\sim$0.29 km/s) are notable characteristics. 
This spectral signature is potentially consistent with outgassing directed away from the observer (e.g., a night-side jet), suggesting a more plausible expansion velocity of $\sim$0.4 km/s.

The beam-averaged column density of $\rm NH_3$ in a given metastable ($J$, $K$=$J$) state can be calculated by the following equation (see details in \citet{1997A&A...325L...5B}):
\begin{equation}
    \langle N(J, K=J) \rangle = 6.8 \times 10^{12} \frac{J + 1}{J}  \int T_{\mathrm{MB}}(v) \, dv\,    (\rm cm^{-2})\,,
\end{equation}
$J$ and $K$ are the rotational quantum numbers.
The integral term represents the total intensity over a velocity range. 
The relationship between the beam-averaged column density and production rate from \citet{1982Icar...51....1S} (see also \citet{1997A&A...325L...5B}) was used to estimate $\rm NH_3$ production rate when the ammonia coma region is much smaller than the FOV:

\begin{equation}
    \langle N(\text{NH}_3) \rangle =  \frac{4\tau Q(\text{NH}_3)}{\pi \Delta^2 \theta^2}
\end{equation}
where $\Delta$ is the geocentric distance and $\theta$ is the half-power beam width of the antenna.
Considering the lifetime of $\rm NH_{3}$ was about $5.56 \times 10^{3}$ s at 1 AU (the average value for a quiet and an active Sun from \citet{2015P&SS..106...11H}) and the small $\rm NH_{3}$ outgassing velocity suggested by the $\rm NH_{3}(3,3)$ line profile (about $0.3\pm 0.1$ km/s from the Doppler shift and FWHM), the ammonia coma region was much smaller than the width of the projected antenna beam.
To roughly estimate the $\rm NH_3$ production rate, we assumed a Boltzmann distribution for the level population $n$($J$,$K$=$J$)=$\langle$$N$($J$,$K$=$J$)$\rangle$/$\langle$$N$($\rm NH_{3}$)$\rangle$ and used the value for a rotational temperature of 60 K, as given by \citet{1997A&A...325L...5B} and used by \citet{1996AAS...188.6212P}.
The production rate and column density are given in Table~\ref{tab7}.

\begin{table}
    \centering
    \caption{The ammonia production rate and column density in 12P/Pons-Brooks}
    \setlength{\tabcolsep}{3mm}
    \begin{threeparttable}
    \begin{tabular}{ccccc}
    \toprule
    lines & \makecell[c]{$\langle N(J,K=J) \rangle$ \\ $[10^{12} \, \rm{cm}^{-2}]$} & \makecell[c]{$\langle N(\rm{NH}_{3}) \rangle$ \\ $[10^{12} \, \rm{cm}^{-2}]$} & \makecell[c]{$Q_{\rm NH_{3}}$ \\ $[\times 10^{27} \, \rm{molec. \cdot s^{-1}]}$} \\
        \midrule
        (3, 3) & \makecell[c]{$<0.41$ \\ $0.20 \pm 0.05$} & \makecell[c]{$<1.70$ \\ $0.83 \pm 0.21$ } & \makecell[c]{$<3.91$ \\ $1.91 \pm 0.48$} \\
        \bottomrule
    \end{tabular}
    \end{threeparttable}
    \label{tab7}
\end{table}

\section{Discussion}
\label{sec4}

\subsection{Evolution of water production rates}

\begin{figure}
   \centering
   \includegraphics[width=\columnwidth]{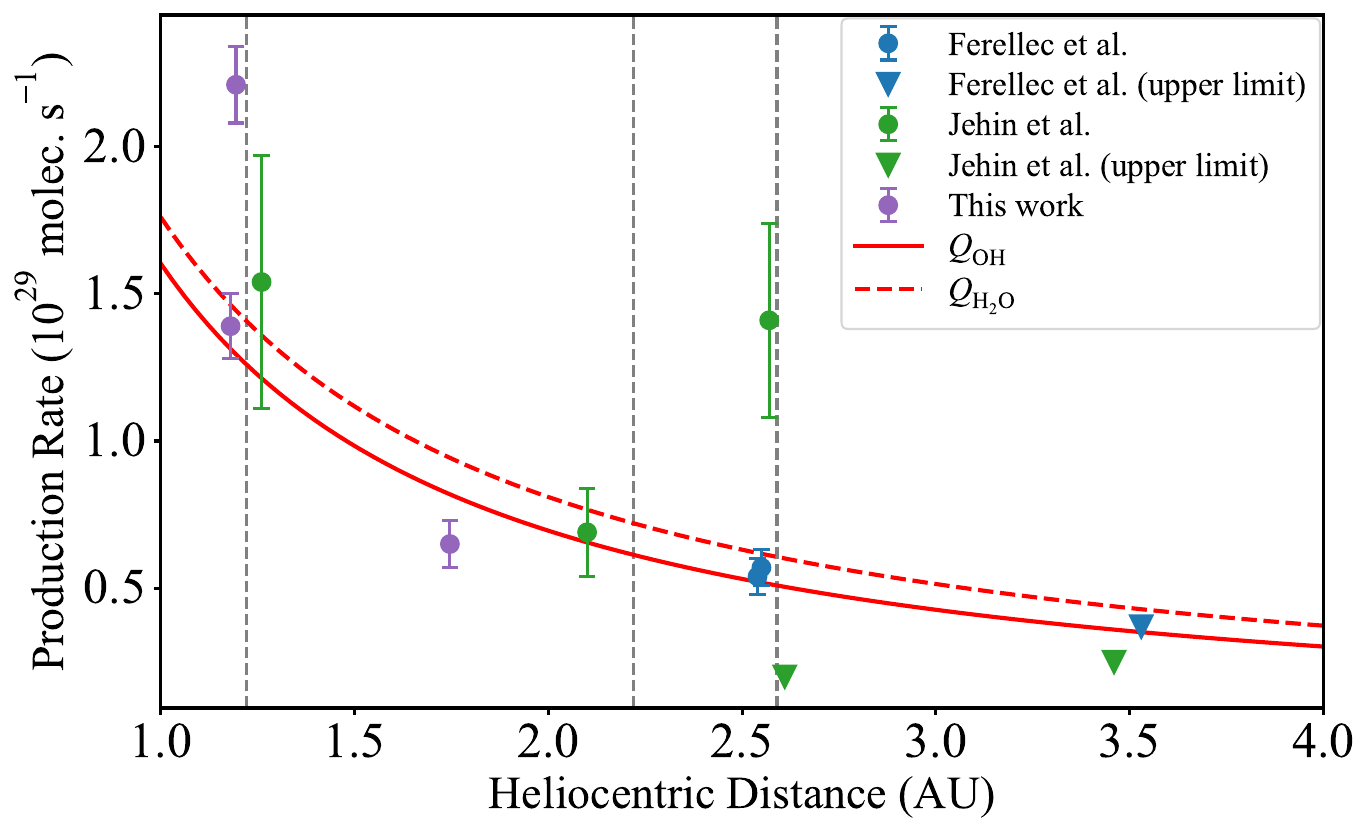}
      \caption{OH production rates of comet 12P/Pons-Brooks as a function of heliocentric distance. The measurements from different sources are shown with different symbols. Error bars represent 1$\sigma$ uncertainties, and triangles indicate upper limits. The red solid line shows the best power-law fit, excluding data immediately after the outbursts on November 14 and February 29 (marked by gray dashed lines). The dashed red line indicates the curve of water production rate inferred by the OH curve.}
         \label{fig4}
   \end{figure}

Fig.~\ref{fig4} illustrates the variation of OH production rates in comet 12P/Pons-Brooks with respect to heliocentric distance.
The data points are compiled from different sources: \citet{2024MNRAS.534.1816F}, \citet{2023ATel16338....1J,2023ATel16282....1J,2024ATel16498....1J,2024ATel16408....1J}, and our current work. 
A power-law fit (shown as a red solid line) was applied to the non-outburst data, yielding the relationship $Q_{\rm OH}=(1.60 \pm 0.16)\,r_{\rm h}^{(-1.20 \pm 0.15)} \times 10^{29}\rm \,molec. \cdot s^{-1}$, corresponding to a water production rate variation $Q_{\rm H_2O}=(1.76 \pm 0.18)\,r_{\rm h}^{(-1.20 \pm 0.15)} \times 10^{29}\rm \,molec. \cdot s^{-1}$ (red dashed line).
The analysis of OH production rate dependence on heliocentric distance reveals significant variations among different dynamical populations of comets, with the power-law exponent serving as a critical parameter for characterizing these variations. 
Our analysis shows that 12P shows a relatively shallow power-law slope of -1.2 based on data from 4.0 AU to 1.0 AU inbound, which is closer to the taxonomic class of depleted long period comets according to \citet{2019Icar..317..610C}.
Dynamical families show a clear progression in power-law steepness: Long-period comets (LPCs; including young, old, and dynamically new comets) exhibit the flattest pre-perihelion exponents with a mean value of -2.32 (range: -7.8 to -0.5), while Jupiter-family comets (JFCs) display much steeper slopes averaging -4.14 (range: -0.4 to -11.9). 
HTCs show intermediate behavior with a steep average exponent of -4.38 according to \citet{1995Icar..118..223A}.
Notably, 12P exhibited exceptionally high activity near perihelion, with its water production rate reaching $10^{30}\rm molec. \cdot s^{-1}$ that results in a steeper power-law index of -3 \citep{2024DPS....5630110C}.
This significant change in its power-law index could be linked to its physical and dynamical properties.

The relatively flat power-law exponent observed for 12P has important implications for understanding its physical properties. 
This behavior might reflect distinctive characteristics in surface morphology, distribution of active regions, or compositional properties of volatile reservoirs within the nucleus. 
The moderate heliocentric dependence suggests that 12P maintains relatively efficient outgassing processes even at larger heliocentric distances compared to typical Halley-type comets. 
Such behavior warrants further investigation to determine whether it represents a transitional evolutionary state between LPCs and HTCs or a unique compositional signature.

\subsection{The active area}
To further investigate the comet’s evolving activity, we calculated the minimum active area corresponding to the water production rates using a sublimation model\footnote{https://ice-sublimation-tool.astro.umd.edu/} \citep{1979M&P....21..155C}.
We set obliquity $=0^\circ$ to maximize the sublimation averaged over the entire surface, hence getting a lower limit of the total active area. 
We further assumed an albedo of 0.04 and an infrared emissivity of 100\% \citep{2020ApJ...893L..48X}. 
The resulting minimum active areas $A_{\rm min}$ in our work were about $99\sim122 \,\rm km^2$ on March 2 during an outburst and decreased to about $61\sim75 \,\rm km^2$ on the next day.
If we use the fitted power-law function results and account for long-term variations during pre-perihelion (excluding outburst periods), the minimum active area of 12P stabilizes at about $70 \,\rm km^2$.

Under the assumption of a fully active surface (active fraction of 100\%), a lower limit for the nuclear radius of 
$r_{\rm min} =\sqrt{A_{\rm min}/4\pi} =2.22\,\rm km$ was derived from the peak active area observed on March 3, 2024, less affected by an outburst.
\citet{2020RNAAS...4..101Y} calculated a larger radius of $17\pm6\,\rm km$, which likely represents an upper limit due to uncorrected contamination from cometary dust and gas emissions.

\subsection{Abundance of Ammonia}
The $\rm NH_{3}/H_{2}O$ abundance ratio in comets measured so far shows a wide range of values, ranging from 0.1 to 2.0\% across different cometary families \citep{2011ApJ...727...91K, 2016ApJ...820...34D}. 
Notably, $\rm NH_{3}$ is among the molecules whose production ratio with water exhibits a correlation with heliocentric distance.
Recent observations of both JFCs and Oort cloud comets have revealed this behavior (figure 7 of \citet{2023A&A...677A.157D}), suggesting that the production mechanism of $\rm NH_{3}$ might be more complex than direct sublimation from the nucleus.
The observed variations could be attributed to different formation conditions in the early solar system \citep{2011ARA&A..49..471M}, heterogeneous distribution within the nucleus \citep{2019AJ....158..128M}, or secondary production mechanisms in the coma \citep{2004come.book..425F}. 
The recent discovery of ammonium salts ($\rm NH_{4}^{+}X^{-}$) on cometary dust grains has revealed a significant reservoir of nitrogen in comets \citep{2020NatAs...4..533A}. 
These semi-volatile compounds, especially $\rm NH_{4}^{+}Cl^{-}$, with sublimation temperatures of 160-230 K, decompose to produce $\rm NH_{3}$ as a primary product \citep{2019JPCA..123.5805H}. 
This mechanism might explain the observed increase in $\rm NH_{3}/H_{2}O$ ratios as comets approach perihelion, as demonstrated through IR spectroscopic observations of multiple comets \citep{2016Icar..278..301D}.
It appears that the detection of ammonia production at large perihelion distances is primarily from the ice in comet nucleus as the temperature of the dust particles is not high enough for salts sublimation.

We used the previously obtained power-law function between OH production rate and heliocentric distance in Section 4.1 to calculate the theoretical water production rate during the $\rm NH_{3}$ observation period. 
Since the comet was in outburst on that day, we doubled the inferred value on the basis of \citet{2024ATel16408....1J}, who report that $Q_{\rm CN}$ increased by a factor of 2 following the outburst. 
The derived $\rm NH_{3}/H_{2}O$ ratio of 12P at 2.22 AU is $(1.36 \pm 0.34) \%$ using the estimated water production rate of $1.4 \times 10^{29} \, \rm molec.\, s^{-1}$.
Because the $\rm NH_3$ data were acquired during an outburst event, the abundance of ammonia is uncertain and is possibly overestimated.

\begin{figure}
    \centering
    \includegraphics[width=\linewidth]{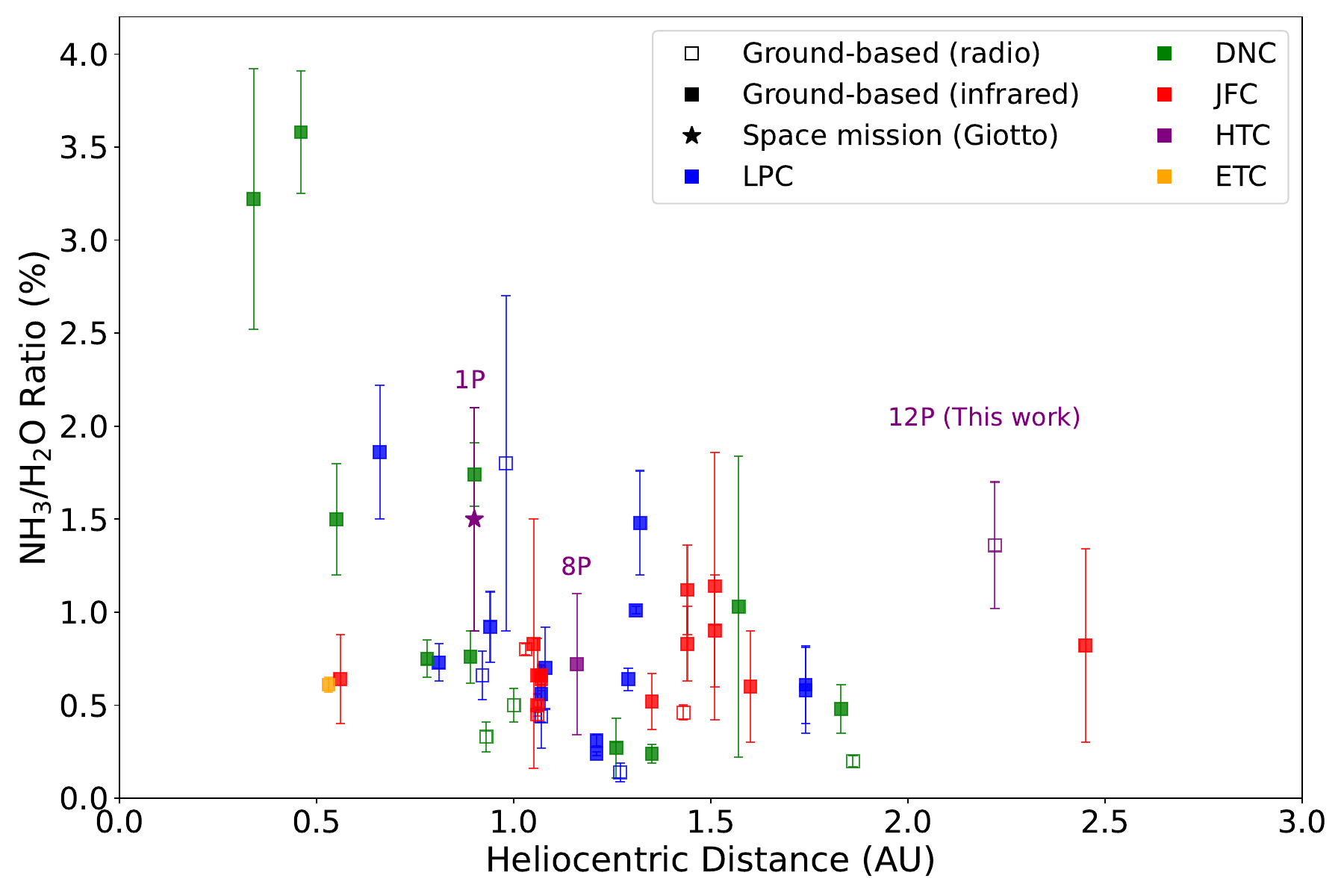}
    \caption{$\rm NH_3/H_2O$ as a function of heliocentric distance updated from \citet{2023A&A...677A.157D}. The measurement in 1P/Halley was acquired by the $Giotto$ space mission. For the dynamical classification of comets in the legend, ETC = Encke-Type Comet, JFC = Jupiter-Family Comet, LPC = Long-Period Comets originating from the Oort Cloud, DNC = Dynamically New Oort Cloud Comet, HTC = Halley-Type Comet.}
    \label{fig:NH3_ratio}
\end{figure}

At similar heliocentric distance, there was only one observation in a Jupiter family comet 17P/Holmes, and the $\rm NH_{3}/H_{2}O$ ratio was $(0.82 \pm 0.52)\%$ at 2.45 AU \citep{2021AJ....162...74L}.
For a HTC like 12P, observations of ammonia in the K-band were even more limited.
\citet{1987cra..proc...85B} reported the K-band observations of ammonia in 1P/Halley with no detection.
However, the $\rm NH_{3}/H_{2}O$ ratio in 1P/Halley was also estimated by the photodissociation products of ammonia (e.g. $\rm NH_{2}$, $\rm NH$) from optical spectra (e.g. $(0.20 \pm 0.14)\%$, \citet{1991ApJ...368..279W}), ultraviolet spectra (0.44-0.94\%, \citet{1993ApJ...404..348F}), $Giotto$ spacecraft ion mass spectrometer (IMS) (1-2\%, \citet{1987A&A...187..502A}) and neutral mass spectrometer (NMS) ($1.5^{+0.5}_{-0.7}\%$, \citet{1994A&A...287..268M}).
The ammonia abundance of another HTC, 8P/Tuttle, was $(0.72 \pm 0.38)\%$ estimated by infrared high-resolution spectroscopy (see Fig.~\ref{fig:NH3_ratio}).
Our derived $\rm NH_{3}/H_{2}O$ ratio places 12P among the higher end of observed ratios for its heliocentric distance, notably exceeding the 8P/Tuttle measurement and comparable to active JFCs such as 9P/Tempel 1 ($1.14 \pm 0.72\%$ at 1.51 AU), 10P/Tempel 2 ($1.12 \pm 0.24\%$ at 1.44 AU) and LPCs such as C/2013 R1 (Lovejoy) ($1.48 \pm 0.28\%$ at 1.32 AU). 
The $\rm NH_{3}/H_{2}O$ ratio measured in 12P beyond 2 AU appears relatively high (Fig.~\ref{fig:NH3_ratio}).
However, as illustrated in the figure, only two points from the same dynamically new comet (DNC) C/2012 S1 (ISON) exhibit elevated abundances at smaller heliocentric distances, whereas other comets show values concentrated below 2\% without a pronounced trend. 
Actually, the observations beyond 1.5 AU are also not numerous enough to obtain a trend.
This ambiguity complicates the assessment of heliocentric dependence and necessitates further observational verification. 
Nevertheless, it should be emphasized that our $\rm NH_{3}$ abundance estimate for 12P lacks contemporaneous water production measurements, which may introduce uncertainties.
Moreover, it is critical to note that detection verification must take precedence before drawing definitive conclusions.

\section{Conclusions}
\label{sec5}

The TMRT was used to observe a Halley-type comet 12P/Pons-Brooks in L-band and K-band before perihelion, focusing on OH radical and $\rm NH_{3}$, 
respectively, during the outburst periods.

   \begin{enumerate}
    \item The expansion velocities of water molecules, inferred from trapezoid fitting of OH line profiles, showed significant temporal variations. 
    Assuming an ejection velocity of OH in the rest frame of $\rm H_2O$, we obtained a lower expansion velocity of 0.5 km/s when the comet was at 1.75 AU. 
    The expansion velocity decreased from 1.4 km/s to 1.0 km/s within two consecutive days in early March of 2024, which was possibly related to the outburst on February 29th.
    \item The OH production rate exhibited variations that closely follow the pattern of expansion velocities. 
    The production rates evolved from $(0.65\pm 0.08)$ $\rm \times 10^{29}$ $\rm molec.\cdot s^{-1}$ on January 20, 2024, peaked at $(2.21 \pm 0.13)$ $\rm \times 10^{29}$ $\rm molec.\cdot s^{-1}$ on March 2, and subsequently decreased to $(1.39 \pm 0.11)$ $\rm \times 10^{29}$ $\rm molec.\cdot s^{-1}$ on March 3.
    \item The pre-perihelion OH production rate demonstrated a clear heliocentric dependence, following a power-law relationship with an index of -1.2 beyond 1 AU, which is flat compared to much steeper water curve immediate before perihelion.
    \item We constrained the minimum active area of 12P to be above $61 \,\rm km^2$, corresponding to a lower limit nuclear radius of 2.22 km.
    \item A 3$\sigma$ detection of the $\rm NH_{3}$ (3,3) inversion line was achieved, characterized by an unusually narrow line width consistent with an expansion velocity of about 0.3 km/s. 
    The derived NH$_3$/H$_2$O ratio of $(1.36 \pm 0.34)\%$ might be overestimated, due to the possible underestimation of water production rate.
   \end{enumerate}

\begin{acknowledgements}
      We sincerely thank Dr. Dominique Bockel{\'e}e-Morvan for providing constructive comments and valuable suggestions, which have greatly improved the quality of this work.
      The authors thank all the staff of the Tianma-65m Radio Telescope at Shanghai Astronomical Observatory for their assistance.
      We thank Dr. Juan Li from Shanghai Astronomical Observatory for helpful suggestions in the noise assessment of observations.
      This work is financially supported by the National Natural Science Foundation of China (Grant Nos.12233003, 12173093 and 12033010), the science research grants from the China Manned Space Project with No. CMSCSST-2021-B08.
      We acknowledge the support of the Minor Planet Foundation of Purple Mountain Observatory, China.
\end{acknowledgements}

\bibliographystyle{aa} 
\bibliography{Yourfile}
\end{document}